\documentclass{article}  
\usepackage{breckenr2005}
\frompage{000} \topage{000}                                              
\def\rightmark{ATLAS Heavy Ions}
\def\leftmark{S.~White}
 
\title{Heavy Ion Physics with the ATLAS Detector } 
\authors{
{Sebastian N. White$^1$ %
}\\[2.812mm]
{\normalsize
\hspace*{-8pt}$^1$ Brookhaven National Lab \\ 
Upton, N.Y., USA 11973\\[0.2ex] 
}}
 
\abstract{Soon after the LHC is commissioned with proton beams the ATLAS experiment will begin studies of Pb-Pb collisions with a center of mass energy of $\sqrt s_{NN}=5.5$ ~TeV. The ATLAS program is a natural extension of measurements at RHIC in a direction that exploits the higher LHC energies and the superb ATLAS calorimeter and tracking coverage. At LHC energies, collisions will be produced with even higher energy density than observed at RHIC. The properties of the resulting hot medium can be studied with higher energy probes, which are more directly interpreted through modification of jet properties emerging from these collisions, for example. Other topics which are enabled by the 30-fold increase in center of mass energy include probing the partonic structure of nuclei with hard photoproduction (in UltraPeripheral collisions) and in p-Pb collisions. Here we report on evaluation of ATLAS capabilities for Heavy Ion Physics.  }

\keyword{LHC,ATLAS, heavy-ion,jet quenching} 
\PACS{25.75.Nq --25.75.Dw}
 
\begin{document}
 
\maketitle
\setcounter{page}{1}

\section{Introduction}\label{intro}
	      The many exciting results presented at this conference and the rapid developments in Heavy ion physics since RHIC came on line 4 years ago clearly show the impact on this field of opportunities opened by new multi-purpose detectors and advances in energy and luminosity. The field has also benefited from the focused activity of 4 simultaneous experiments at RHIC. Competition is clearly a good thing!
	      
	      Before the end of the decade the present research at RHIC will evolve in 2 different directions.
	      The RHIC detectors -STAR and PHENIX are planning detector upgrades to exploit an increase in accelerator luminosity available in the RHIC-II phase\cite{Harris}. Essentially at the same time the 2 large general purpose detectors CMS and ATLAS will come on line at CERN complementing the ALICE approach to Heavy Ions. 
	      
	      The ATLAS group has been evaluating the opportunities in Heavy Ion physics for several years. The collaboration as a whole presented a letter of intent\cite{loi} to the CERN LHC committee a year ago (Spring '04) detailing the results of performance studies based on detector simulations and received encouragement to proceed with this program. In the past year we have continued these studies after concluding that there is clear opportunity with essentially no modification to the detector (except for the addition of Zero Degree Calorimeters for trigggering and event characterization).
	      
		We have focused on aspects of the CERN program which capitalize on higher available energy and the the unique features of ATLAS- high quality calorimetry with the highest segmentation (both longitudinal and transverse) and sensitivity of the LHC detectors, large rapidity coverage and its external muon spectrometer. We summarize several of these
topics below.

\section{Simulations}\label{techno}  

	The main detector systems in ATLAS are
\begin{itemize}
\item{the inner tracker- consisting of 3 silicon pixel layers, 8 layers of silicon strip and 39 layers of Transition Radiation tracking(TRT) which cover the pseudorapidity range $|\eta|\leq 2.5$. Because of the high occupancy of the TRT in central Pb-Pb collisions we neglect, for now, TRT hits when evaluating tracking performance.}
\item{the electromagnetic and hadronic calorimeters which cover the interval $|\eta|\leq 4.9$. The calorimeter features very high segmentation with $\sim 200 k$~EM channels and 4 layers of depth segmentation. The main EM compartment has towers of $(0.025)^2$ in $\delta\eta\times\delta\phi$ and a pre-shower segmentation of $\delta\eta=0.003$.}
\item{the muon system is an independent spectrometer outside the calorimeter volume and covers
$|\eta|\leq 2.7$. Energy loss in the calorimeter limits muon measurements to $p_{lab}^{\mu}\geq5$ GeV, approximately independent of $\eta$. Since average hadron $p_T$ is lower than in pp collisions
the muon spectrometer is cleaner in Pb-Pb running than in the pp program for which it was designed.}
\end{itemize}
 
\subsection{Simulation tools}\label{details}

	The detector performance is analyzed using a Geant3 based simulation including the full digitization chain. ATLAS has extensive experience in comparing this simulation to testbeam performance. Heavy ion events were simulated in different data sets (of collision centrality ranging from 
b=0 to 15 fermi) using the generator HIJING 1.38. In this model the maximum charged particle multiplicity reaches
$\frac{dN}{d\eta}\sim 3,200$ whereas a naive extrapolation of RHIC multiplicities vs cms energy yields
$\frac{dN}{d\eta}\sim 1,300$ at LHC energy, so we consider these simulations to be a very stringent test of detector performance.
	For studies of jet-reconstruction and dimuon analysis we merged Pythia objects with the underlying  Heavy Ion event and passed them through the same Geant simulation. In currently ongoing studies we introduce other aspects of heavy ion events- such as elliptic flow.
		 
\begin{figure}[htb]
\vspace*{-.8cm}
                 \insertplot{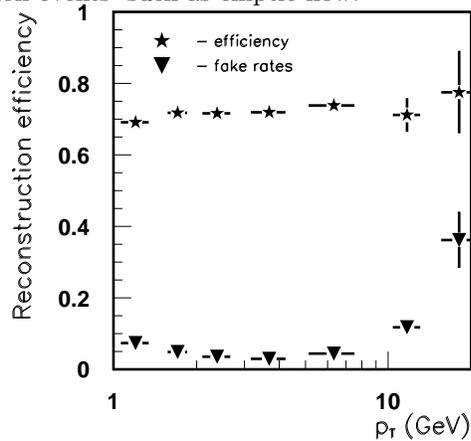}
\vspace*{-2cm}
\caption[]{Track reconstruction.}
\label{fig1}
\end{figure}

\subsection{Simulation results-event characterization}\label{characterization}
	Analysis of Global properties of Heavy Ion events and the choice of  minimum bias trigger are closely related. The goal of the Heavy ion program is to characterize events according to centrality
	(effectively the collision impact parameter, b) and reaction plane and study the evolution of properties of jets, for example, as centrality and orientation with the reaction plane are varied. Although most global observables are strongly correlated with b the real figure of merit is the range of impact parameters covered by the trigger and the analyzing power of the detectors over this full range. This is necessary to cover the full transition from the most peripheral (pp like) events to the most central. One measure of jet suppression used at RHIC (so-called $R_{CP}$: for central-over-peripheral) directly compares the 2 extreme event classes. For these reasons and because they are the only way to access the physics of Ultraperipheral Collisions we plan to add Zero Degree Calorimeters in the far forward region of the ATLAS intersection region.
	
		Nevertheless our simulations of the existing detector have shown that measurements of centrality over most of the range of interest  are accurate to $\delta b \sim 1$~fermi.
		
\subsection{Simulation-tracking}\label{track}
	In the most central collisions the occupancy of the silicon tracker is below 2$\%$ in the pixel layers and ranges from 10-20$\%$ in the strip layers. Figure \ref{fig1} shows that under these circumstances the standard ATLAS tracking algorithm gives ~$70\%$ reconstruction over the $p_T$ range of 1-10 GeV/c.
	We have required 10 out of 11 hits along the track (13 in the "end-cap region") and that at most 1 hit is shared per track. We also find a momentum resolution which ranges from 2-3$\%$ (depending on $\eta$) - limited by multiple scattering. The fake rate is typically $\leq5\%$ and could be reduced at the higher momenta by comparing with energy in the corresponding calorimeter cell.
	
\subsection{Simulation-jet reconstruction}\label{jets}
	Two aspects of jet physics are unique to Heavy ions and therefore hadn't received earlier attention in simulations of the calorimeter performance. 

\begin{figure}[htb]
\vspace*{-.8cm}
                 \insertplot{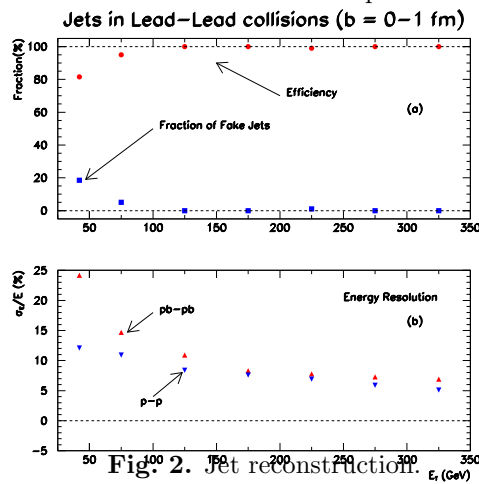}
\vspace*{-2cm}
\caption[]{Jet reconstruction.}
\label{fig2}
\end{figure}

	We plan to study the modification of jet shapes and fragmentation in the highest centrality collisions. Calculations\cite{Gyulassy} of jet quenching at LHC energy predict that the jet fragmentation spectrum is suppressed ($R_{AA}$($p_T$)) by a factor which ranges from 10 to 2 over the $p_T$ range of 10 to 100 GeV/c. Furthermore the properties of jet quenching are expected to differ significantly for jets arising from b or c-quark fragmentation when compared to gluon or lighter quark fragmentation. For these reasons we would like to study jet reconstruction and energy resolution in events with the highest underlying activity and down to as low a 
	jet energy as possible. We also plan to use the muon spectrometers to identify b-jets using a soft lepton tag as well as detached vertices from b-decay.
					 
\begin{figure}[htb]
\vspace*{-.8cm}
                 \insertplot{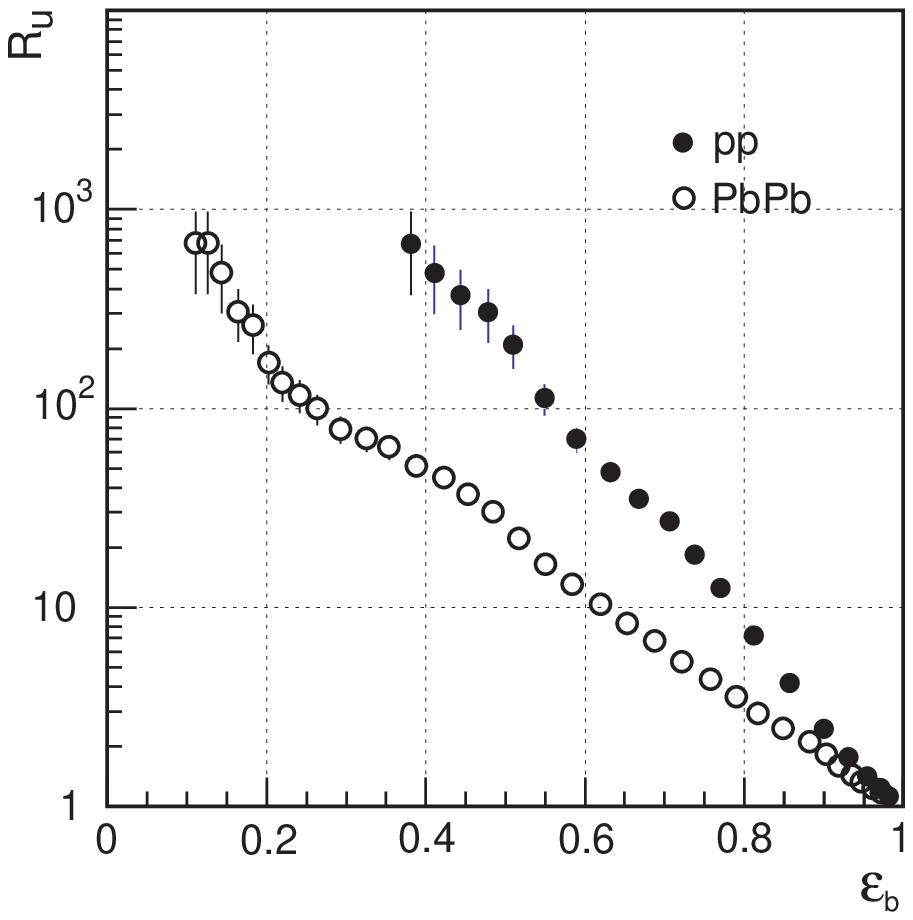}
\vspace*{-2cm}
\caption[]{b-jet tagging efficiency from secondary vertex.}
\label{fig3}
\end{figure}
		
	The other aspect of heavy Ion physics which is unique (but with possible implications also for the ATLAS pp diffractive program) is the extension of jet studies to low $p_T$ in diffractive (Ultraperipheral ) ion collisions. In this case the main consideration is jet finding at low $p_T$ (=10-20 GeV) but with relatively little underlying event activity. Then the main consideration will be finding jets and measuring their energy when charged particles are dispersed by the 2 Tesla magnetic field of the ATLAS spectrometer.
			
		For the jet quenching studies we have used the standard ATLAS sliding window algorithm with $\delta\eta\times\delta\phi=0.4\times 0.4$ and a two-step background subtraction procedure. We are still actively studying tools for jet reconstruction in Heavy Ion events so the initial results shown in Figure \ref{fig2} may be regarded as a current status of the effort. At this stage the minimum jet energy which can reliably be reconstructed in the highest multiplicity events is $\sim 30-40$~GeV since the typical energy in a random jet cone is about 50~GeV. At higher jet energies the energy resolution (also shown in Figure \ref{fig2}) approaches the pp performance as expected.
					 
			We studied b-jet tagging by examining the reconstruction efficiency and rejection power  (against light quark jets) using one of the tools available for b-tagging in ATLAS- secondary vertex finding ( we will extend these studies using also a soft muon candidate collinear with the jet axis). The results of the  simulation are shown in Figure \ref{fig3}  where we have superimposed on Pb-Pb collisions a standard Pythia object from the ATLAS Higgs particle search - $pp\rightarrow WH\rightarrow l\nu b\bar b$ and $l\nu u\bar u$. With this tool the tagging efficiency is lower than in pp collisions but it should improve significantly with the soft lepton tag.

\subsection{Simulation- Upsilon }\label{Muon}
	Heavy quarkonia suppression is a key element of the ATLAS heavy ion program and our results on 
$\Upsilon$ recostruction are discussed extensively in the ATLAS Letter of Intent\cite{loi}. The number of 
$\Upsilon\rightarrow\mu^+\mu^-$ events recorded in a typical 1 month run (the expected Heavy Ion running period for 1 year of normal LHC operation) is $\geq 10^4$ and the $\Upsilon$ mass resolution 
ranges from 110 MeV to 150 MeV depending on the range of pseudorapidity. This should be adequate for resolving the $\Upsilon$ states and a detailed study of their suppression with centrality.

\section{Photoproduction Studies in Ultraperipheral collisions}\label{UPC}
	At RHIC the study of high energy photoproduction in Heavy Ion collisions is starting to yield results\cite{Dorgali}. Whereas the equivalent photon spectrum in $\sqrt s_{NN}=200$~GeV Au-Au collisions corresponds roughly to fixed target $\gamma$-A collisions at the Fermilab tagged photon lab,
the reach at LHC will be several orders of magnitude higher. This means that in addition to measurements diffractive vector meson production ATLAS will study deep inelastic photoproduction of dijets, heavy flavors, etc. The program resembles that of HERA but with nuclear targets and will probe the partonic structure of nuclei to $x_{parton}\sim 10^{-4}$. Again the above topics including low $p_T$ jet measurement and heavy flavor tagging are relevant but in this case the underlying event is relatively quiet- $\gamma$-A events have very low multiplicity.

\begin{figure}[htb]
\vspace*{-.8cm}
                 \insertplot{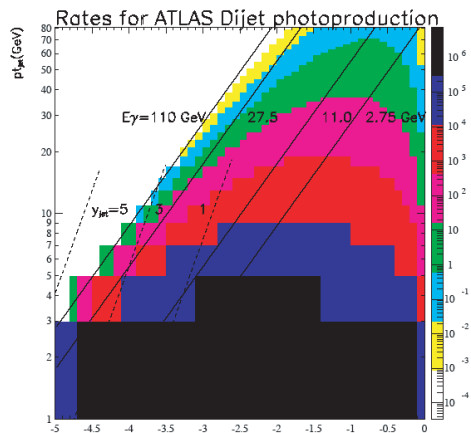}
\vspace*{-2cm}
\caption[]{Event yields in a 1 month ATLAS run.}
\label{fig4}
\end{figure}

	We have started to evaluate ATLAS capabilities for di-jet photoproduction using\cite{vogt} available Pb structure functions and the Weizsacker-Williams spectra modified for tagging by nuclear breakup\cite{baur}. Initial results are shown in Figure \ref{fig4}. Here we display event yields for a 1 month
run for the process $\gamma+Pb\rightarrow jet+jet+X$ per bin of 2 GeV in jet $p_T$ and 0.25 in $\frac{\delta x_{parton}}{x_{parton}}$. One sees that the event yields are substantial and the mean jet rapidity falls well within the pseudrapidity $\leq 4.9$ coverage of the ATLAS calorimeter. 
	At RHIC\cite{Dorgali} we learned that requiring a neutron tag (using the forward Zero Degree Calorimeter(ZDC)) and a rapidity gap (an $\eta$ interval with less than 2-4 GeV of calorimeter activity in the ATLAS case) provides a very powerful trigger and event selection for studying this class of event. The ATLAS collaboration has recently completed the design and performance simulation of the ZDC which should be operating in time for LHC startup.
	
\section{Conclusion}
	The ATLAS collaboration is well advanced in preparations for a broad program of Heavy Ion studies which will begin in 2008-2009. We look forward to future "Nuclear Dynamics" workshops where results from LHC and RHIC-II will be presented.
	
\section{Acknowledgement}
	I would like to thank the organizers of this workshop and particularly Dr. Bauer for a very stimulating and enjoyable meeting.This work was supported in part under
DOE Contract number DE-AC02-98CH10886.

\vfill\eject
\end{document}